\documentclass[11pt]{article}

\usepackage[dvips]{graphicx}
\title{\vspace*{-2.0cm}
\begin{flushright}
\normalsize{ FERMILAB-Pub-04/003-T\\EFI-03-45}
\end{flushright}
\vspace*{2.5cm}\textbf{Visible Sector Supersymmetry Breaking}\\
\textbf{Revisited}
\vspace*{1.2cm}}
\author{\large\textbf{Piyush Kumar$^{\dag}$}
\textbf{and} \textbf{Joseph D. Lykken$^{\dag,*}$}\\ \\[0.5cm]
$^{\dag}$\normalsize\emph{Enrico Fermi Institute and Department of
Physics}\\ \emph{University of Chicago, Chicago, IL 60637, USA}\\ \\
$^*$\normalsize\emph{Fermi National Accelerator Laboratory}\\
\emph{P.O. Box 500, Batavia, IL 60510, USA}}

\begin{document}

\maketitle
\vspace*{1.6cm}

\begin{abstract}
We revisit the possibility of
``visible sector'' SUSY models: models which are
straightforward renormalizable extensions of the Minimal
Supersymmetric Standard Model (MSSM), where SUSY is broken
at tree level. Models of this type were abandoned twenty
years ago due to phenomenological problems, which we review.
We then demonstrate that it is possible to construct simple
phenomenologically viable visible sector SUSY models. Such
models are indeed very constrained, and have some inelegant features.
They also have interesting and distinctive phenomenology.
Our models predict light
gauginos and very heavy squarks and sleptons. The squarks
and sleptons may not be observable at the LHC. The LSP is
a stable very light gravitino with a significant Higgsino
admixture. The NLSP is mostly Bino. The Higgs boson is naturally
heavy. Proton decay is sufficently
and naturally suppressed, even for a cutoff scale as low as $10^8$ GeV.
The lightest particle of the O'Raifeartaigh sector (the LOP) is stable,
and is an interesting cold dark matter candidate.
\end{abstract}

\thispagestyle{empty}
\newpage


\section{Introduction}

Supersymmetry (SUSY) is a beautiful idea which starts to lose some
of its lustre
when one tries to build complete models that break SUSY
in a realistic way. This lack of felicity has been accentuated in recent
years by experimental constraints that seem to force some
degree of unexplained tuning in any viable model \cite{Kane:2002ap}.

In all popular models, supersymmetry breaking occurs in a
``hidden sector" and supersymmetry breaking is ``mediated" to the
visible sector by indirect interactions.
The known scenarios for the mediation of
SUSY breaking in hidden sector models
were classified in \cite{gordyreview} as
{\it gravity mediation}, {\it gauge
mediation}, and {\it bulk mediation}.
Simply put, in gravity mediation
the soft parameters arise due to couplings which are Planck suppressed,
{\it i.e.} they vanish as
$M_P \rightarrow \infty$. In gauge mediation, the soft parameters
arise from loop diagrams involving new messenger fields with
Standard Model (SM) quantum numbers.
In bulk mediation, the hidden and observable
sectors reside on different branes separated in extra dimensions,
and SUSY breaking is mediated by fields which propagate
in the bulk.

In this paper we revisit the possibility of what one
might call ``visible sector'' SUSY models: models which are
straightforward renormalizable extensions of the Minimal
Supersymmetric Standard Model (MSSM), where SUSY is broken
at tree level. Models of this type were abandoned twenty
years ago due to phenomenological problems which we will review
in the next section. In addition, hidden sector models seemed
more attractive theoretically, as they have a natural tie-in to
the grand unification and Planck scales, and exhibit radiative
electroweak symmetry breaking due to the large top quark Yukawa
coupling.

In this paper we demonstrate that it is possible to construct
phenomenologically viable visible sector SUSY models. Such
models are indeed very constrained, and have some inelegant features.
They also have interesting and distinctive phenomenology.

Our model (it is really a class of models) possesses an extra
low energy $U(1)$ gauge group, under
which the two Higgs fields are charged with the same sign. This
implies that the $\mu$ term is forbidden in our model, but an effective
$\mu$ term is generated by the spontaneous breaking of the extra $U(1)$.
In addition, as will be shown, adding an extra $U(1)$ also helps
to sufficiently suppress the $B$ and $L$ violating interactions.

Our model can be considered as a complete effective field theory
description of physics below a cutoff scale which can be as low as
about $10^8$ GeV. As such, it is impressively simple.
It is also similar to the effective models proposed by
Banks \cite{Banks:2002ar}, which are conjectured to mock up the
effects of cosmological supersymmetry breaking by string effects.

The paper is organized as follows. In Section $2$, earlier models
of low energy supersymmetry are reviewed and some of their
problems are discussed. In Sections $3$ and $4$, we introduce the model,
outline its main features and calculate its spectrum. In Section
$5$, we comment on phenomenological implications of the model.
Technical details are provided in the Appendix.

\section{Earlier Attempts}

Any supersymmetric model limited to the Standard Model gauge group
has two immediate problems:
\begin{itemize}
\item Renormalizable, and thus unsuppressed, $B$ \& $L$ nonconserving
interactions are present.

\item There is a mass sum rule at tree level which is
phenomenologically untenable, because it leads to very light
superpartners which have not been observed.
\end{itemize}
In the Minimal Supersymmetric Standard Model (MSSM), these
problems are dealt with in the following way \cite{pa} :
\begin{itemize}

\item A discrete symmetry (R-parity) is imposed on the Lagrangian
which forbids all $B$ \& $L$ violating renormalizable
interactions.

\item Supersymmetry is assumed to be explicitly broken by soft
(super-renormalizable) terms which allow us to give acceptable
values to particle masses. These soft terms are put in \emph{by
hand}.
\end{itemize}
However, it is still possible to have dimension five R-parity
conserving but $B$ \& $L$ violating interactions such as
\((QQQL)_{F}\) and \((UUDE)_{F}\), which among other things seem
likely to induce proton decay at a rate greater than the current
bounds set by SuperKamiokande \cite{pz}. Thus the above remedies
are not complete, in addition to being {\it ad hoc}.

The approach adopted in this paper is to suppose that the gauge
group which survives down to ordinary energies contains an
additional factor $G$. As was shown in \cite{pg}, $G$ must contain
a $U(1)$ factor, and so the simplest choice is just $U(1)$. Now,
if all the quark and lepton superfields have a $U(1)$ charge
with the \emph{same} sign, then all \(d=4\) R-parity violating
and \(d=5\) R-parity conserving
interactions involving only quarks and leptons are forbidden.

Giving all the quarks and
leptons $U(1)$ charges of the same sign has another advantage. It
leads to the possibility that all squarks and sleptons can be
sufficiently massive. In addition, the $\mu$ term is
forbidden, while an ``effective" $\mu$ term is generated when the field
coupled to $H_u$ and $H_d$ obtains a vacuum expectation value.
This is important for providing sufficient masses to the
charginos, as will be seen later.

However, the situation is more subtle. Adding an extra gauge group
introduces new anomalies in the MSSM, which was originally anomaly
free. So extra fields must be added to cancel these anomalies.
Constructing an anomaly free model with a viable phenomenology is
not easy. To appreciate these problems better, let us look at them
in greater detail.

In any $\mathcal{N}=1$ supersymmetric gauge theory with gauge
group \(G=\prod_\alpha G^{\alpha}\), there is an interaction
between vector superfields \(V^{\alpha}\) and chiral superfields
\({\Phi}_a\). The scalar, spinor and vector components of
\(V^{\alpha}\) are $D^{\alpha}$, ${\lambda}^{\alpha}$ and
$V_{\mu}^{\alpha}$ while the scalar and spinor components of
\({\Phi}_a\) are $F_a$, ${\phi}_a$ and ${\psi}_a$. The tree
level effective potential for the scalar components \({\phi}_a\)
of \({\Phi}_a\) is given by:

\begin{eqnarray}
V(\phi) &=& \frac{1}{2} \sum_{\alpha} (D^{\alpha})^2 +
\sum_{a} |F_{ai}|^2 \nonumber\\
&=& \sum_{\alpha} \frac{1}{2} g_{\alpha}^2(\sum_{a}
{\phi}_{ai}^{\dag} T^{{\alpha}a}_{ij}
{\phi}_{aj}+{\xi}^{\alpha})^{2} + \sum_{a} |\frac{\partial
W}{\partial {\phi}_{ai}}|^{2} \; ,
\end{eqnarray}

\noindent where \(\alpha\) labels different factors of the gauge
group, \(a\) the various chiral superfield representations and
\(i\) the components of each representation. The gauge couplings
for \(G^{\alpha}\) are \(g_{\alpha}\), and \(T^{\alpha a}\) are the
generators of \(G^{\alpha}\) in the representation of
\({\phi}_{a}\). \(W\) denotes the superpotential. The
Fayet-Iliopoulos (FI) couplings  ${\xi}^{\alpha}$
are only present if there is a U(1) factor
\cite{ph}. The spontaneous breaking of supersymmetry at tree
level leads to a mass relation \cite{pi} :

\begin{equation}
\sum_{J} (-1)^{2J}(2J+1)\; \mathrm{Tr}\,(m_J^2) = \sum_{\alpha}
\frac{1}{2}\;g_{\alpha}^2 \langle D^{\alpha} \rangle\;
\mathrm{Tr}\,(t^{\alpha})\; ,
\end{equation}

\noindent where $m_J$ is the mass matrix for spin $J$ fields,
\(\alpha\) now runs only over $U(1)$ factors and
\begin{equation}
\langle D^{\alpha} \rangle = (\sum_{a} {\langle {\phi}_{ai}
\rangle }^{\dag} T^{\alpha a}_{ij} \langle {\phi}_{aj} \rangle
+{\xi}^{\alpha})\; .
\end{equation}

\noindent In our model we will introduce a single extra $U(1)$
factor, which we will denote by $\tilde{U}(1)$. Now observe
that if \(\langle \tilde{D} \rangle \neq 0\)
and the trace over quarks and leptons is separately nonzero, as is
the case when all the squarks and sleptons are given
\(\tilde{U}(1)\) charges of the same sign, then there is a
possibility that all the sparticles can be made to receive large
masses.

It has proven difficult to construct a
renormalizable and anomaly free model of this type. Previous attempts
at models along these lines has led to unacceptable features such as
the presence of a color breaking minimum, the absence of a $\tilde{D}$ term
vev, or both. The best developed earlier models appear in
\cite{pj}, \cite{pk}. However in \cite{pj},
$\mathrm{Tr}\,(\tilde{U}(1)) \neq 0$,
so there is a $\tilde{U}(1)$-gravitational anomaly and
a quadratically divergent renormalization of the FI term \cite{pl},
while in
\cite{pk}, the problem of finding the global
minimum has not been correctly dealt with.

A more successful recent example are the models of Cheng, Dobrescu
and Matchev \cite{Cheng:1998nb}. These are completely chiral
renormalizable models with an anomaly free extra $U(1)$.
These models have in addition a hidden sector with extra gauge
interactions that generate $F$ term SUSY breaking dynamically,
as in standard gauge mediation, and spontaneously break the extra
$U(1)$ at a scale $\sim 10^3$ TeV. The dimensionless parameters of
the models are tuned at tree level such that the $D$ term vev ends
up of order 100 GeV. These models are phenomenologically viable and
have some attractive theoretical features.

Our model also provides a totally anomaly free solution to the problems
outlined at the beginning of this section. We
have a single input scale of order 20 TeV.
Supersymmetry and the extra $U(1)$ are broken spontaneously at
tree level. As in the models of \cite{Cheng:1998nb}, a single tree
level tuning is necessary to generate the electroweak scale from
the 20 TeV input scale, but it is not a fine tuning since the
radiative corrections are suppressed.
$B$ \& $L$ violating interactions are greatly suppressed.

\section{The Model}

Our model is built along the lines of \cite{pg}, \cite{pj} and
\cite{pk}.
A table of all the chiral superfields,
together with their quantum numbers, is provided in the Appendix.
The model has several exotics with Standard Model quantum numbers:
a color octet chiral superfield
\(K\), an \(SU(2)\) triplet superfield \(T\), and two vectorlike
pairs of hypercharged chiral superfields $J_i$, $J^c_i$, $i$$=$$1,2$.
These fields
are introduced to cancel the \(SU(3)^{2}\tilde{U}(1)\),
\(SU(2)^{2}\tilde{U}(1)\), and $U(1)^2\tilde{U}(1)$
anomalies. To construct a
completely anomaly free model, several MSSM singlet fields
which are only charged under \(\tilde{U}(1)\) are also needed.

Supersymmetry is broken at tree level by an O'Raifeartaigh sector \cite{OR},
generating $F$ term vevs. The model has
Fayet-Iliopoulos terms for both the $\tilde{U}(1)$ and hypercharge,
leading to $D$ term vevs. Electroweak symmetry and the $\tilde{U}(1)$
gauge symmetry are broken at tree level.
In order to generate one-loop gaugino masses that are large enough to
satisfy current experimental bounds, the $F$ and $\tilde{D}$ term vevs
must be of order 20 TeV. Thus to obtain the proper electroweak scale,
we require a single tree level tuning of the Fayet-Iliopoulos
parameters. This is the
least attractive property of visible sector SUSY
breaking. However the tree level tuning is robust against radiative
corrections, {\it i.e.} masses which are of order the electroweak scale
at tree level remain of order the electroweak scale after radiative
corrections.

The superpotential of the model is
\begin{eqnarray}
W = W_{\rm Yukawa} + W_{\rm O'R} + W_{\mu} + \tilde{W} \; .
\end{eqnarray}
This superpotential consists of four pieces. The first piece is
the superpotential of the MSSM without a $\mu$ term:
\begin{eqnarray}
W_{\rm Yukawa} = y_{E}L\bar{E}\mathcal{H} + y_{D}Q
\bar{D}\mathcal{H} + y_{U}Q\bar{U}\mathcal{H'} \; .
\end{eqnarray}
The second piece is an O'Raifeartaigh sector which breaks SUSY
and the $\tilde{U}(1)$
spontaneously at tree level. Due to this $F$ term breaking, all of
the MSSM gauginos, squarks and sleptons receive soft-breaking masses,
either at tree level, or at one-loop, or both.
\begin{eqnarray}
W_{\rm O'R} &=
\displaystyle{{\lambda}_{K}K^2X_{1}+{\lambda}_{T}T^2X_1
+\sum_{i=1}^{2}{\lambda}_{J}J_{i}J_{i}^{c}X_{1}+
 {\lambda}_{R}\,\sum_{i=1}^{11}R_i^{2}X_{1}} \\
&+(fY+M_2)X_1X_2-f{\mu}^2Y+M_1X_2X_3 \; . \nonumber
\end{eqnarray}
The third piece consists of the MSSM Higgs fields and some
MSSM singlets, some of which are charged under $\tilde{U}(1)$.
This sector spontaneously also breaks the $\tilde{U}(1)$ gauge symmetry,
and simultaneously generates an effective $\mu$ term.
\begin{eqnarray}
W_{\mu} = \frac{1}{2}m{\phi}^2+({\mu}'+\bar{g}\phi)SN
-{\kappa}^2\phi+\frac{1}{3}\beta{\phi}^3+\bar{M} N\,P
-g_{H}S\mathcal{H} \mathcal{H'} \; .
\end{eqnarray}
The last piece couples the hypercharged exotics $J_i$, $J^c_i$
to the MSSM singlet field $P$ that appears in $W_{\mu}$:
\begin{eqnarray}
\tilde{W}= {\lambda}_{P}J_i\,J_i^c\,P
\; .
\end{eqnarray}
This additional coupling is needed to explicitly break
an accidental global $U(1)$ symmetry otherwise present
in the model, which is spontaneously broken when some of the fields
get vevs.

It is important to note that there is some freedom in the
choice of $W_{\mu}$. One could extend the Higgs sector in
several ways consistent with cancellation of anomalies.
For example, it would be interesting to incorporate the ``$\mu$-less''
models of Nelson {\it et al} \cite{Nelson:2002ca}.

The scalar potential generated is :
{\setlength\arraycolsep{2pt}
\begin{eqnarray}
V &=& |\frac{\partial W}{\partial {\phi}}|^2 +
\frac{1}{2}{[D_{SU(3)}]^2 + \frac{1}{2}[D_{SU(2)}]}^2+
\frac{1}{2}\;g_{y}^{2}\;[\frac{1}{6}Q^{\dag}Q-\frac{2}{3}{\bar{U}}^{\dag}
\bar{U}+\frac{1}{3}{\bar{D}}^{\dag}{\bar{D}}-
\frac{1}{2}L^{\dag}L+{\bar{E}}^{\dag}{\bar{E}}+
\frac{1}{2}{\cal{H'}}^{\dag}{\cal{H'}}-\nonumber \\
& &\frac{1}{2}{\cal{H}}^{\dag}{\cal{H}}+\sum_{i=1}^{2}
J_{i}^{\dag}J_{i}-\sum_{i=1}^{2}J_{i}^{c\dag}J_{i}^{c}+\xi\;]^{2}+\frac{1}{2}\;{\tilde{g}}^{2}\;[Q^{\dag}Q+{\bar{U}}^{\dag}
\bar{U}+{\bar{D}}^{\dag}{\bar{D}}+
L^{\dag}L+{\bar{E}}^{\dag}{\bar{E}}-2{\cal{H}}^{\dag}{\cal{H}}-\nonumber\\
& &2{\cal{H'}}^{\dag}{\cal{H'}}
-2K^{\dag}K-2T^{\dag}T-2\sum_{i=1}^{2}J_{i}^{\dag}J_{i}-2\sum_{i=1}^{2}J_{i}^{c\dag}J_{i}^{c}+
4X_1^{\dag}X_1-4X_2^{\dag}X_2
+4X_3^{\dag}X_3+\nonumber \\
& &4S^{\dag}S-4N^{\dag}N+4P^{\dag}P-2\sum_{i=1}^{11}R_i^{\dag}R_i
+\sum_{i=1}^{3}O_{i}^{\dag}O_{i}
+4V^{\dag}V
+\tilde{\xi}\;]^{2} \;
.\label{eq:one}
\end{eqnarray}}

\noindent It is straightforward to show the following features of
the above model:

\begin{itemize}
\item \(Y,\,\phi,\,X_1,\,X_2, S\, \&\, N \) get vevs of order 20 TeV.
\item The exotics $K$, $T$, and $J_i$, $J^c_i$ all get masses of order 20 TeV.
\item For \(2g_H^2<g^2\), the Higgs doublets
can be brought down to a form \({\mathcal{H}}=
\left(\begin{array}{c}{h_d}\\{0}\end{array} \right) \) and
\({\mathcal{H'}}= \left(\begin{array}{c}{0}\\{h_u}\end{array}
\right) \) at the potential minimum, corresponding to the
conservation of electric charge.
\end{itemize}

\noindent For a suitable range of parameters, the global minimum
can be found by solving the following equations for $h_u$ and
$h_d$:
{\setlength\arraycolsep{2pt}
\begin{eqnarray}
&\hskip-26pt
(g_H^2+4\tilde{g}^2-\frac{g^2+g_y^2}{4})h_d^2
+(\frac{g^2+g_y^2}{4}+4\tilde{g}^2)h_u^2-8\tilde{g}^2(V_1^2-V_2^2+S^2-N^2
+\frac{1}{4}\tilde{\xi})
+g_H^2\,S^2+\frac{1}{2}g_y^2\xi
= 0\,,\label{eq:six} \nonumber\\
&\hskip-26pt
(g_H^2+4\tilde{g}^2-\frac{g^2+g_y^2}{4})h_u^2
+(\frac{g^2+g_y^2}{4}+4\tilde{g}^2)h_d^2-8\tilde{g}^2(V_1^2-V_2^2+S^2-N^2
+\frac{1}{4}\tilde{\xi})
+g_H^2\,S^2-\frac{1}{2}g_y^2\xi = 0\,,\label{eq:seven}
\end{eqnarray}}

\noindent and the following equations for $X_1$, $X_2$, $S$ and
$N$:

{\setlength\arraycolsep{2pt}
\begin{eqnarray}
M_1^2\, V_1 V_2&=&f^2\,({\mu}^2-V_1V_2) (V_1^2+V_2^2) \;,
\label{eq:two}\nonumber\\
\frac{M_1^2V_2^2}{V_1^2+V_2^2}&=&[16 \tilde{g}^2
(V_1^2-V_2^2+S^2-N^2)+ 4
\tilde{g}^2(\tilde{\xi}-2h^2)]\label{eq:three}
 \; ,\nonumber\\
({\bar{g}}^2\,N^2+4{\tilde{g}}^2\langle \tilde{D} \rangle +
g_H^2\, h^2) \langle S \rangle &=& -\bar{g} \langle N^* \rangle
(\beta{\phi}^2+m\,
\phi- {\kappa}^2)\; ,\label{eq:a} \nonumber\\
({\bar{g}}^2\,S^2-4{\tilde{g}}^2\langle \tilde{D} \rangle +
{\bar{M}}^2) \langle N \rangle &=& -\bar{g} \langle S^* \rangle
(\beta{\phi}^2+m\, \phi- {\kappa}^2)\; ,\label{eq:b}
\end{eqnarray}}
\noindent where $\langle X_1 \rangle = V_1$,  $\langle X_2 \rangle
= V_2$, $\langle S \rangle = S$,  $\langle N \rangle = N$, $h^2 =
h_u^2 + h_d^2 = $(174 GeV)$^2$. The vev of $Y$ is given by $
\langle Y \rangle =-\langle M_2 \rangle /f$ and $\langle \phi
\rangle =-{\mu}'/\bar{g}$.

The above vacuum structure ensures that
charge and color are not broken. The derivation of these
equations and various constraints is carried out in the Appendix.
As will be seen later,
the qualitative nature of the vacuum will not change
even after radiative corrections are taken into account.

As already explained, this model has a single input scale on the
order of 20 TeV, and a single tree level tuning to get the
proper scale of electroweak symmetry breaking. The precise values
of the input parameters are not very much constrained.
As an example, an appropriate choice of parameters compatible
with all constraints is:

{\setlength\arraycolsep{2pt}
\begin{eqnarray}
f&=&1;\;{\lambda}_{J}={\lambda}_{K}={\lambda}_{T}={\lambda}_{P}=0.75;\;
\beta=0.5;\;{\lambda}_{R}=0.25 \; ,
 \nonumber \\ M_1&=&M_2=\bar{M}=18\;\mathrm{TeV};\;\mu=23.7\;\mathrm{TeV};
\;m \simeq 20\;\mathrm{TeV} \; ,\nonumber \\
V_1&=&V_2=S={\mu}'=20\;\mathrm{TeV};\;N \simeq 20\;\mathrm{TeV};\;
\kappa \simeq 19 \;\mathrm{TeV}\;, \nonumber
\\
\tilde{g}\tilde{\xi} &\simeq& (20.1\;\mathrm{TeV})^2 \;\;
\bar{g}=1;\;g_H=0.45;\;\tilde{g}=0.1. \label{eq:eight}
\end{eqnarray}}

\noindent where we have also shown all of the resulting vevs.
The electroweak scale is set by a tree level tuning of $\tilde{\xi}$.
For the above choice of parameters, we have:

\begin{eqnarray}
h_u^2 =
\frac{8{\tilde{g}}^2(V_1^2-V_2^2+S^2-N^2+{\tilde{\xi}}/4)-g_H^2S^2}
{g_H^2+8{\tilde{g}}^2}
- \frac{g_y^2\; \xi}{g^2+g_y^2-2g_H^2}\; , \\
h_d^2 =
\frac{8{\tilde{g}}^2(V_1^2-V_2^2+S^2-N^2+{\tilde{\xi}}/4)-g_H^2S^2}
{g_H^2+8{\tilde{g}}^2} + \frac{g_y^2\; \xi}{g^2+g_y^2-2g_H^2}\; .
\end{eqnarray}

\noindent The hypercharge $FI$ parameter $\xi$ determines
$\mathrm{tan}(\beta) = h_u/h_d $ and, together with $V_1, V_2,
S\;\&\;N$, it also determines the
the $Z$ boson - $B'$ boson mixing. For {\it e.g.} a $\tan\beta$
of 2, we need:

\begin{equation}
g_y\,\xi \simeq -(0.06\;\mathrm{TeV})^2\; .
\end{equation}

\subsection{$R$ symmetry}

One of the problems
of earlier models of tree level supersymmetry breaking was the presence of a
spontaneously broken
continuous $R$ symmetry \cite{pc},\cite{pd},\cite{pg}.
This is a consequence of the result shown in \cite{pm}, that for
any generic and calculable theory the existence of an $R$
symmetry is a necessary condition, while the existence of a
spontaneously broken $R$ symmetry is a sufficient condition for
supersymmetry breaking. A spontaneously broken $R$ symmetry leads
to the existence of a massless Goldstone boson, which
is undesirable for phenomenology \cite{pv}.

Our model does not have the above problem because it has no
continuous $R$ symmetry.
Supersymmetry, however, is still broken spontaneously.
This is compatible with the result in \cite{pm}, because the
superpotential of our model is nongeneric, {\it i.e.} it does not have
all possible terms consistent with the symmetries of the theory.
Terms like $K^2X_3$, $Y^2$, etc. which are otherwise allowed in
the superpotential  are not present in our model. Their omission
is thus not natural in the usual sense. However,
non-renormalization theorems assure us that if these terms are not
present in the tree level lagrangian, they are not generated to
any order in perturbation theory. The model is natural in this
weaker sense.

\subsection{Discrete symmetries}

Our superpotential has several discrete symmetries, a $Z_4$ discrete
symmetry which we call O'charge, and several $Z_2$ parities.
The O'charges of the various superfields are given in the table in the
Appendix. The O'charges $q_i$, where $i$ runs over all the chiral superfields,
satisfy the relations
\begin{eqnarray}
\sum q_i^3 = 0 \,{\rm \ mod\ }4 \; , \nonumber\\
\sum q_i = 0 \,{\rm \ mod\ }4 \; .
\end{eqnarray}
This means that O'charge can be considered as an anomaly free
discrete gauge symmetry \cite{Ibanez:1991pr,Banks:1991xj},
which is thus respected
also by higher dimension operators that could contribute to the
superpotential. We will assume by the same token that the $Z_2$
parities are merely accidental, and thus are not respected
by higher dimension operators that contribute to the
superpotential.

At the same time that the $\tilde{U}(1)$ symmetry is spontaneously broken, the $Z_4$
O'charge invariance is broken down to a residual $Z_2$ parity. We call the remaining
unbroken discrete gauge symmetry O'parity.

\section{Superpartner Spectrum}

\subsection{Fermions}

The mass terms for the gauginos, higgsinos and other chiral
fermions receive contributions from two sources at tree level --
one due to the superpotential, and the other due to trilinear
couplings between the gauginos, fermions and scalars. The
Higgsinos, electroweak gauginos and various O'Raifeartaigh sector
fields which are only charged under $\tilde{U}(1)$ mix due to the
effects of electroweak and $\tilde{U}(1)$ symmetry breaking. The
charged higgsinos, charged winos and charged O'Raifeartaigh sector
fields combine to form mass eigenstates with charge $\pm 1$ called
charginos. Similarly, the neutral higgsinos, neutral winos and
neutral O'Raifeartaigh sector fields mix to form mass eigenstates
called neutralinos. The gluinos on the other hand, being color
octet fermions, receive no contributions to their masses at tree
level. At one loop, diagonal mass parameters are induced for all
the gauginos. This is the same as in ordinary models of gauge
mediated supersymmetry breaking. The contribution to the masses of
the gluinos, winos, bino and b'ino at one loop at the scale $\mu$
is given by an equation similar to that in \cite{pn}:
\begin{equation}
M_a(\mu)=\frac{\alpha_a(\mu)}{4\pi} \sum_{i} {\Lambda}(i)\,
n_a(i)\; ,
\end{equation}

\noindent where $a$ labels the gauge group $SU(3)$, $SU(2)$,
$U(1)_Y$ and $\tilde{U}(1)$, $i$ runs over all chiral fields and
$n_a(i)$ is one-half the Dynkin index of the $a^{th}$ gauge group.
The difference from the equation in \cite{pn} is that the mass
scale $\Lambda$ is different for different chiral fields. For our
choice of parameters, $ \sum_i {\Lambda}(i)\,n_a(i) \approx
3.72\,V_1$ for $ a=SU(3)$, $\approx 2.48\,V_1$ for $a=SU(2)$,
$\approx 1.8\,V_1$ for $a=U(1)$ and $\approx 40.5\,V_1$ for
$a=\tilde{U}(1)$. This results in a one-loop gluino mass of
approximately 600 GeV.

\begin{figure}
\centerline{
          \includegraphics{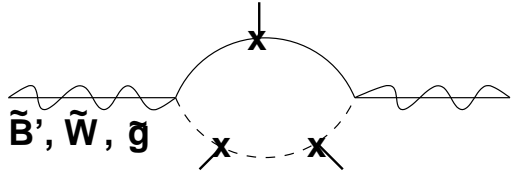}}
\caption{Contributions to gaugino masses at one loop}
\end{figure}

\noindent In the gauge eigenstate basis, the chargino mass terms
in the Lagrangian (at tree level) are :

{\setlength\arraycolsep{2pt}
\begin{eqnarray}
-\triangle \mathcal{L}&=&
\frac{1}{2}{(\Psi^{\pm})}^{T}\mathcal{M}_{c}(\Psi)^{\pm}
\\
&=& g_H\langle S \rangle \tilde{H}_{u}^{+} \tilde{H}_{d}^{-}+g
\langle \mathcal{H}^{0*}_{d} \rangle
\tilde{W}^{+}\tilde{H}_{d}^{-}+ g \langle \mathcal{H}^{0*}_{u}
\rangle \tilde{W}^{-}\tilde{H}_{u}^{+}+\sum_{i=1}^2 {\lambda}_{J}
\langle X_1 \rangle{\psi}_{J_i}{\psi}_{J_i^c}+ 2{\lambda}_{T}
\langle X_1 \rangle{\psi}_{T^+}{\psi}_{T^-} \,.\nonumber
\end{eqnarray}}

\noindent To find the correct chargino content, we also need to
add the one-loop mass for the charged wino and diagonalize the
resulting matrix. The mass terms for the O'Raifeartaigh sector
fields are decoupled from the MSSM fields. This leads to the
following mass squared eigenvalues:

{\setlength\arraycolsep{2pt}
\begin{eqnarray}
M^2_{J_i}=M^2_{J_i^c}&=&({\lambda}_J\,V_1)^2=(15\;\mathrm{TeV})^2;\;
M^2_{T^+}=M^2_{T^-}=(2{\lambda}_T\,V_1)^2=(30\;\mathrm{TeV})^2;\nonumber\\
M^2_{\tilde{C}_1,\tilde{C}_4}&=&{(9\;\mathrm{TeV})}^2;\;
M^2_{\tilde{C}_2,\tilde{C}_3}={(130\;\mathrm{GeV})}^2.
\end{eqnarray}}

\noindent for a reasonable $\tan\beta$ value of 2.

\noindent The neutralino mass terms can be analyzed in a similar
way. In the gauge eigenstate basis, the neutralino mass terms (at
tree level) are:

{\setlength\arraycolsep{2pt}
\begin{eqnarray}
-\triangle \mathcal{L} &=&
\frac{1}{2}{(\Psi^0)}^{T}\mathcal{M}_{N}{(\Psi^0)}\nonumber \\
&=& g_H\langle S \rangle \tilde{H}_u^0 \tilde{H}_d^0+g_H \langle
H_u^0 \rangle \tilde{X_4}\tilde{H}_d^0+g_H \langle H_d^0 \rangle
\tilde{X_4}\tilde{H}_u^0+ \frac{g'}{\sqrt{2}} \langle H_u^{0*}
\rangle \tilde{B}\tilde{H}_{u}^{0}-\frac{g'}{\sqrt{2}} \langle
H_d^{0*} \rangle \tilde{B}\tilde{H}_{d}^{0}+\nonumber\\
& &\hskip-20pt \frac{g}{\sqrt{2}} \langle H_d^{0*} \rangle
\tilde{W}^0\tilde{H}_{d}^{0}-\frac{g}{\sqrt{2}} \langle H_u^{0*}
\rangle \tilde{W}^0\tilde{H}_{u}^{0}-2\sqrt{2}\tilde{g} \langle
H_u^{0*} \rangle \tilde{B'}\tilde{H}_{u}^{0}-2\sqrt{2}\tilde{g}
\langle H_d^{0*} \rangle
\tilde{B'}\tilde{H}_{d}^{0}+4\sqrt{2}\tilde{g} \langle {X_1^*}
\rangle \tilde{B'}\tilde{X_1}-\nonumber \\
& &\hskip-20pt 4\sqrt{2}\tilde{g}\langle {X_2^*} \rangle
\tilde{B'}\tilde{X_2}+M_1\tilde{X_2}\tilde{X_3}
+f
\langle X_2 \rangle \tilde{Y}\tilde{X_1}+ f \langle X_1 \rangle
\tilde{Y}\tilde{X_2}
+\lambda_{R}\langle X_1
\rangle \sum_{i=1}^{11}{\psi}_{R_i}{\psi}_{R_i}+\nonumber\\
& &\hskip-20pt \bar{M}\,\tilde{N}\tilde{P}+(m+2\beta \langle \phi
\rangle)\,\tilde{\phi}\tilde{\phi}+\bar{g} \langle N \rangle
\tilde{S} \tilde{\phi}+\bar{g} \langle S \rangle \tilde{N}
\tilde{\phi}+2\,\lambda_{T}\langle X_1 \rangle
{\psi}_{T^0}{\psi}_{T^0}+ 2\,\lambda_{K}\langle X_1
\rangle{\psi}_{K}^{(N)}{\psi}_{K}^{(N)} \;.
\end{eqnarray}}

\noindent It is clear that the mass terms for
${\psi}_{R},{\psi}_{T^0}$ and neutral components of ${\psi}_{K}$
are decoupled from other terms. Therefore, it is sufficient to
diagonalize the remaining thirteen dimensional mass matrix. There
are two zero eigenvalues at tree level. One of them corresponds to
the goldstino ($\tilde{G}$), which is \textit{exactly} massless as
is expected for a spontaneously broken globally supersymmetric
theory. This can be seen easily, because the components of
$\tilde{G}$ are of the form $(\langle D_{\alpha} \rangle / \sqrt{2}$,
$\langle F_i \rangle )$
and it is also annihilated by $\mathcal{M}_{N}$. The other
massless field corresponds to the photino. However, it is massless
only at tree level. Again, as in the case of charginos, we need to
take the mass parameters for the b'ino, bino and neutral wino
induced at one loop into account. Diagonalizing the resulting mass
matrix gives the following mass eigenvalues for our choice of
parameters:

{\setlength\arraycolsep{2pt}
\begin{eqnarray}
M^2_{R_i}=(10\;\mathrm{TeV})^2;\;M^2_{T^0}&=&M^2_{K^{(N)}}=
(30\;\mathrm{TeV})^2;\nonumber\\
M^2_{\tilde{N}_1,\tilde{N}_{13}} \simeq (30\;\mathrm{TeV})^2;\;
M^2_{\tilde{N}_2,\tilde{N}_{12}} &\simeq&
(30\;\mathrm{TeV})^2;\; M^2_{\tilde{N}_3,\tilde{N}_{11}} \simeq
(25\;\mathrm{TeV})^2;\nonumber\\
M^2_{\tilde{N}_4,\tilde{N}_{10}} \simeq (10\;\mathrm{TeV})^2;\;
M^2_{\tilde{N}_5,\tilde{N}_9} &\simeq& (9\;\mathrm{TeV})^2;\;
M^2_{\tilde{N}_6}=(132\;\mathrm{GeV})^2;\nonumber
\\M^2_{\tilde{N}_7}=(32\;
\mathrm{GeV})^2;& &\;M^2_{\tilde{N}_8}=0\; .
\end{eqnarray}}

\noindent A massless goldstino is expected even after the
introduction of a non-gauge invariant mass parameter for the
gauginos in the mass matrix, because it is a non-perturbative
result. Moreover, since the radiative corrections are small, the
field content of the goldstino stays pretty much the same, with
approximately the following components:

{\setlength\arraycolsep{2pt}
\begin{equation}
\tilde{G} \simeq -0.46 \tilde{B'}-2.2\times 10^{-3}
\tilde{H_d}^0-1.1\times 10^{-3} \tilde{H_u}^0-0.57 \tilde{P}+0.26
\tilde{\phi} -0.57\tilde{X_3}+0.26 \tilde{Y} \; .\nonumber
\end{equation}}
\noindent Thus, the goldstino in our model is mostly made of
O'Raifeartaigh sector fields. However, it has a small but
interesting higgsino content, which might be relevant for
phenomenology \cite{Gogoladze:2002xp}. We also expect a light
NLSP, which in our case is mostly a Bino. The exact values are not
very crucial, and can be made higher by changing the parameters
slightly.

\subsection{Scalars}

When supersymmetry is spontaneously broken at tree level, the scalar
mass degeneracy is lifted by $D$ terms.
Since charge and color are unbroken, the only
non-zero $\langle D \rangle$ terms are for $y$ of $U(1)$,
$\tilde{y}$ of $\tilde{U}(1)$ and $t_3$ of $SU(2)$. $\langle
\tilde{D} \rangle$ provides a large contribution at tree level,
contributing to the mass squared of the squarks and sleptons of
the order of the input scale - $(20 \; \mathrm{TeV})^2 $, as seen
in the Appendix. The squarks and sleptons also receive
contribution to their masses by radiative corrections. At one
loop, the contributions to their masses arise from graphs in
Figure 2.

\begin{figure}
\begin{center}
\includegraphics[height=1.0in, width=1.5in]{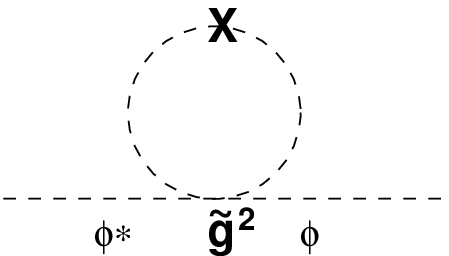}
\hspace{1.5in}
\includegraphics[height=1.0in, width=2.0in]{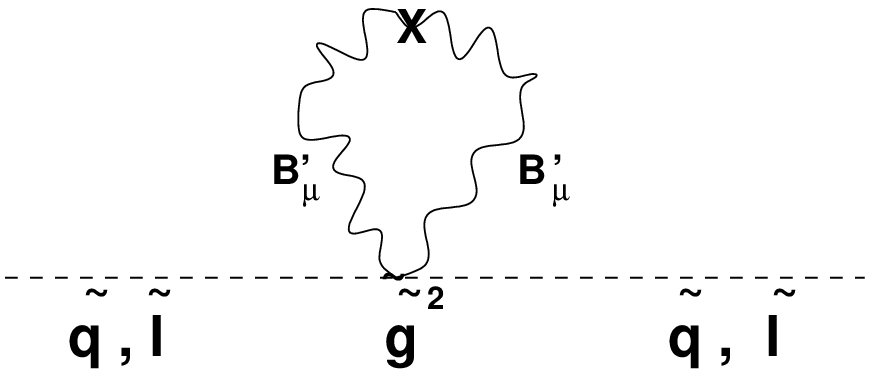}
\end{center}
\caption{Contributions to squark and slepton masses at one loop}
\end{figure}

The dominant contribution to both graphs comes only from the
$\tilde{U}(1)$ gauge group. The values of the graphs are given by:

\begin{eqnarray}
(\Delta M)^2_a \simeq \frac{\tilde{g}^2}{16{\pi}^2}(\sum_{i}
\tilde{q}_{i} M^2_{{\phi}_i});\qquad (\Delta M)^2_b \simeq
\frac{\tilde{g}^2}{16{\pi}^2}(h^2+4(V_1^2+V_2^2+S^2+N^2))\; .
\end{eqnarray}
where $\tilde{q}_i$ are the $\tilde{U}(1)$ charges of the scalar
fields and $i$ runs over all the scalar fields in our model. For
our model, the first graph gives a negative contribution while the
second graph a positive one. For our choice of parameters, as in
(\ref{eq:eight}), $(\Delta M)^2_a \simeq \,-1.5*10^{-3}\,V_1^2$
and $(\Delta M)^2_b \simeq \, 8.1*10^{-5}\,V_1^2$, yielding
$(\Delta M)^2_{one\,loop} \simeq -1.4*10^{-3}\,V_1^2$.

\noindent As can be seen from the scalar potential (\ref{eq:one}),
effective mass parameters for the neutral Higgs fields are
generated at tree level. These effective mass parameters also
receive the above corrections at one loop. The fact that the one
loop contribution is much less than the tree level contribution
lends support to the statement that the tree level equations for $
h_u$ and $h_d$ are robust and
are \textit{not} modified qualitatively even after loop
corrections are taken into account.

\noindent Therefore, the net mass of the squarks and  sleptons
\textit{to} one loop is given by:

\begin{eqnarray}
M_{sq/sl}^2&=&M^2_{tree\,level}+(\Delta M)^2_{one\,loop}\nonumber\\
& \simeq & \tilde{g}^2\,\langle \tilde{D} \rangle + (\Delta
M)^2_{one\,loop} \simeq 0.09 \,V_1^2 \simeq (6\,\mathrm{TeV})^2 \;.
\end{eqnarray}

\noindent We see that the squarks and sleptons in our model are
quite heavy, around 6 TeV. There are further corrections to the
squark and slepton masses from two loop graphs, which are
negligible. In addition, RG evolution has to be used to run these
contributions down to the electroweak scale. However, this does
not affect the qualitative result: squarks and sleptons in the
model are quite heavy. One attractive feature of the above
mechanism is that flavor changing (FCNC) processes are naturally
suppressed, similar to that in gauge mediated models.

\subsection{Gauge Bosons}

The $Z$ boson, $W$ Bosons and the $\tilde{U}(1)$ gauge boson
become massive due to electroweak and $\tilde{U}(1)$ symmetry
breaking. Since $ V_1=V_2=20$ TeV,
the $\tilde{U}(1)$ gauge boson $ B'_{\mu}$ is much heavier than
the $W$ and $Z$ bosons. Apart from the usual mass terms for the $
W^+_{\mu},W^-_{\mu},Z_{\mu}$ and $B'_{\mu}$, there is also a
$Z_{\mu}-B'_{\mu}$ mixing term. The $B'_{\mu}$ mass term and
$Z_{\mu}-B'_{\mu}$ mixing term are given by:

\begin{eqnarray}
M^2_{B'}=8\tilde{g}^2(h^2+4(V_1^2+V_2^2+S^2+N^2));\;\;
M^2_{Z-B'}=2\tilde{g}\sqrt{g^2+g_y^2}(h_u^2-h_d^2)\; .
\end{eqnarray}

\noindent Therefore, the mixing angle defined by
\begin{equation} \label{eq:eleven}
\alpha_{Z-B'}=\frac{1}{2}\;\tan^{-1}(\frac{2M^2_{Z-B'}}{M^2_{B'}-M^2_{Z}}) \;,
\end{equation}
is about $5\times 10^{-6}$, which is well below the experimental
upper bound of $\sim 3\times 10^{-3}$ \cite{po}.

\section{Phenomenology}

\subsection{Gravitino (Goldstino) phenomenology}
One of the distinctive features
of all models of low energy supersymmetry breaking is that the
gravitino, the spin 3/2 superpartner of the graviton, is the LSP.
This is easy to understand. Supersymmetry has to be promoted to a
local symmetry to take gravity into account. So, supersymmetry is
now broken by the super-Higgs mechanism, where the gravitino
acquires a mass by eating the goldstino. The mass of the gravitino
is \cite{pr}:

\begin{equation}
m_{3/2} \sim \frac{\sqrt{{\langle F \rangle}^2+{\langle D
\rangle}^2}}{\sqrt{3}M_p}
\end{equation}

\noindent where $\sqrt{{\langle F \rangle}^2+{\langle D
\rangle}^2}$ is essentially the supersymmetry breaking scale.
Thus, for small $\sqrt{{\langle F \rangle}^2+{\langle D
\rangle}^2}$, as in our model, the gravitino is definitely the
LSP. The mass of the gravitino in our model is $\sim
0.03\,\mathrm{eV}$. The gravitino, by absorbing the goldstino,
inherits its non-gravitational interactions and so can play an
important role in collider physics. Our gravitinos are
sufficiently heavy not to be excluded by current collider limits
\cite{Affolder:2000ef,Brignole:1998me,Dicus:1996ua}.

Since the gravitino is the LSP, we
expect supersymmetric particles, which can be pair produced at
$e^{+}e^-$ colliders through tree-level processes, to decay into
the NLSP (next-to-lightest supersymmetric particle), which then
decays into the gravitino (goldstino). The usual way of analyzing
goldstino interactions is by the method of the effective
lagrangian. From the supercurrent conservation equation, we get
\cite{ps} :

{\setlength\arraycolsep{2pt}
\begin{eqnarray}
{\partial}_{\mu} J^{\mu}_{\alpha} &=& i(\sqrt{{\langle F
\rangle}^2+{\langle D \rangle}^2})
({\sigma}^{\mu}{\partial}_{\mu}\tilde{G}^{\dag})_{\alpha}+{\partial}_{\mu}
j^{\mu}_{\alpha}+...=0 \label{eq:fifteen}\\
j^{\mu}_{\alpha} &\cong& ({\sigma}^{\nu} {\bar{\sigma}}^{\mu}
{\psi}_i)_{\alpha}({\partial}_{\nu}{\phi}^{* i})-\frac{1}{2\sqrt2}{\sigma}^{\nu}
{\bar{\sigma}}^{\rho} {\sigma}^{\mu} {\lambda}^{\dag a} F^a_{\nu
\rho}\label{eq:sixteen}
\end{eqnarray}}

\noindent Here, the ellipses represent contributions which are
unimportant at low energies. Equation (\ref{eq:fifteen}) can be
thought of as the goldstino equation of motion, which can be
derived from the following effective lagrangian:

\begin{equation}
{\mathcal{L}}_{eff}=-i
{\tilde{G}}^{\dag}{\bar{\sigma}}^{\mu}{\partial}_{\mu}\tilde{G}-\frac{1}{(\sqrt{{\langle
F \rangle}^2+{\langle D \rangle}^2})}
(\tilde{G}{\partial}_{\mu}j^{\mu}+h.c.)
\end{equation}
 \noindent Since the
above equation only depends on supercurrent conservation, it does
not depend on the details of supersymmetry breaking. From
(\ref{eq:sixteen}), we see that there are scalar-chiral
fermion-goldstino and gauge boson-gaugino-goldstino vertices,
which can lead to decays to the goldstino (Figure 3).
\begin{figure}
\centerline{\includegraphics[width=9cm]{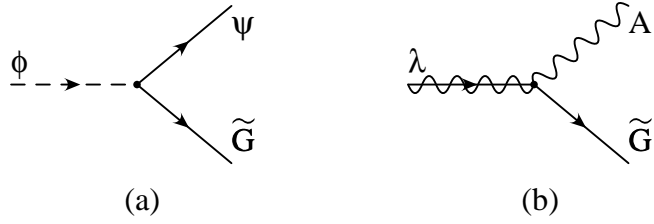}}
\caption{Goldstino interactions with superpartner pairs
(from \cite{ps})}
\end{figure}

\noindent It is important to note that these goldstino
interactions are nonrenormalizble because of $(\sqrt{{\langle F
\rangle}^2+{\langle D \rangle}^2})$  in the denominator. At very
low energies however, these should give the dominant contribution.
The NLSP in our model is mostly a Bino:

\begin{equation}
{\chi}^0\,(NLSP) \simeq -0.9999
 \tilde{B} - 0.003 \tilde{W}^0 +
0.004 \tilde{H_d}^0 - 0.002 \tilde{H_u}^0
\end{equation}

\noindent The dominant decay mode of the NLSP is into the photino
and the gravitino. Its decay rate can be calculated as \cite{pt}:

\begin{equation}
\Gamma (\tilde{\gamma}\rightarrow \gamma \tilde{G}) \simeq
\frac{m_{\tilde{\gamma}}^5\,{\kappa}_{\gamma}}{16 \pi ({\langle F
\rangle}^2+{\langle D \rangle}^2)}
\end{equation}

\noindent where ${\kappa}_{\gamma}$ is the photino content of the
NLSP. In our model, ${\kappa}_{\gamma} \sim 0.15$. The mean decay
length of the NLSP with energy $E$ in the lab frame is therefore,

\begin{equation}
L \sim \frac{1}{{\kappa}_{\gamma}}
(\frac{100\,\mathrm{GeV}}{m_{\tilde{\gamma}}})^5\;(\frac{({{\langle
F \rangle}^2+{\langle D \rangle}^2})^{ 1/4}
}{100\,\mathrm{TeV}})^4\;
(\frac{E^2}{m^2_{\tilde{\gamma}}}-1)^{1/2}\,\times 10^{-2}\; \mathrm{cm}
\end{equation}

\noindent which for our model is around $0.1\;\mathrm{\mu m}$. The
experimental signature for a process like $ e^{+}\,e^- \rightarrow
\tilde{\gamma}\,\tilde{\gamma}$ is thus given by missing
transverse energy, imbalance in the final-state momenta and a pair
of photons. In this case, it is also possible to extend the search
to a portion of the parameter space inaccessible in the
corresponding gravity mediated scenario, where the LSP is
invisible \cite{pu}.

If gravitinos are in thermal equilibrium at
early times and freeze out at temperature $T_f$, their
contribution to the present energy density is \cite{py}:

\begin{equation}
{\Omega}_{3/2}h^2=\frac{m_{3/2}}{keV}\,[\frac{100}{g_*(T_f)}]
\end{equation}

\noindent where $h$ is the Hubble's constant in units of $100
\,\mathrm{km \,sec^{-1}\, Mpc^{-1}}$ and $g_*(T_f)$ is the
effective number of degrees of freedom at $T_f$. Therefore, for
$m_{3/2} <$ keV, as is the case in our model,the gravitinos do
not overclose the universe and late entropy production is not
required.

\subsection{Proton decay and gauge coupling unification}
O'charge and $\tilde{U}(1)$
conservation greatly suppresses $B$ \& $L$ violating interactions.
Proton decay requires dimension 7 operators in the superpotential
of the form
$QQQL\mathcal{H}\mathcal{H'}$
etc., which are suppressed by the cube of the
cutoff. Current lower bounds on the proton lifetime only
constrain the cutoff scale to be greater than about $10^8$ GeV.

The problem of classifying $\tilde{U}(1)$ extensions of the MSSM
which solve the $\mu$ problem, adequately suppress $B \; \& \; L$
violating operators, and keep
gauge coupling unification intact, has been solved
by Erler \cite{da}. In addition, the solutions have been constrained to
respect chirality, so that fields are protected from acquiring
large masses; as well as SU(5) type charge quantization, so that
there are no fractionally charged states. The solutions are based
on the assumption there is a mechanism built in to solve the $\mu$
problem. However, the nature of the mechanism is not specified.
The solutions have the feature that at least two MSSM singlets
charged only under $U(1)$ develop a vev.

It can be seen that one of the $33$ completely chiral solutions of
\cite{da} closely resembles our solution. The slight difference is
due to the fact that our model is not chiral. This is clear,
however, because in our model, the supersymmetry breaking is
effected by an O'Raifeartaigh type model, which is intrinsically
non-chiral. Discounting the fields which are present only to make
the O'Raifeartaigh model work, our model has the satisfying feature
that it also contains two MSSM singlets which develop vevs.
Therefore, it is reasonable to expect that our model leads to
gauge coupling unification.

\subsection{CP Violation} In addition, our model has
implications for $CP$ violation. In our model, there is a
possibility of solving the strong $CP$ problem through the
Nelson-Barr mechanism \cite{pw}. If the determinant of the mass
matrix of all colored fermions is real, ${\Theta}_{QCD}$ vanishes
at tree level. In a spontaneously broken supersymmetric theory,
${\Theta}_{QCD}$ receives corrections proportional to
supersymmetry breaking effects. So, for models of low energy
supersymmetry breaking, ${\Theta}_{QCD}$ does not receive large
corrections \cite{px} and the $CP$ problem can be solved.

\subsection{Cold Dark Matter}

Our model provides a cold dark matter candidate: $R$, the lightest
particle with odd O'parity. For our choice of parameters, the
mass squared of the fields $R_i$ is given by:
\begin{equation}
M_{R_i}^2 \simeq (4\;\mathrm{TeV})^2 \; ,
\end{equation}
\noindent These masses are quite adjustable, so we will exploit this
freedom to assume that one of the $R_i$, called $R$,
is much lighter than the others.
Then the other $R_i$ will decay to this one via higher dimension terms
in the superpotential like $SSR_iRRR$. Note that $R$
is the lightest among all the O'Raifeartaigh
sector particles. We call it the LOP.

The $R$ particle is a Standard Model singlet. The heavy exotics
$K$, $T$, and $J_i$, $J^c_i$ all decay to $R$ particles via
dimension 7 operators like $X_1KRQ\bar{U}\mathcal{H'}$, which
result in 3-body decays like $K\rightarrow R + t + \bar{t}$.
Even though they are from dimension 7 operators, the decay lifetimes
are much less than the age of the universe. This is because the
rates are greatly enhanced relative to the proton decay rate, due
to phase space and the absence of light Yukawa suppression factors.

In the early universe, $R$ scalars will annihilate pairwise into
pairs of Higgs bosons, via a renormalizable $\tilde{D}$ term induced quartic
coupling. Since the annihilation cross section is rather large,
and since the mass of the $R$ particle is fairly adjustable in the
model, the LOP should provide a viable candidate for cold dark matter.

\section{ Summary and Conclusions}

We have presented a class of viable
visible sector SUSY models which do not seem to violate
any current experimental constraints. Such a model could
be a complete and correct description of particle physics
below a cutoff scale as low as $10^8$ GeV.

To summarize the phenomenology, these models predict light
gauginos and very heavy squarks and sleptons. The squarks
and sleptons may not be observable at the LHC. The LSP is
a stable very light gravitino with a significant Higgsino
admixture. The NLSP is mostly Bino. The Higgs boson is naturally
heavy (probably heavier than the MSSM upper bound), but we
have not computed it. The Higgs quartic coupling will have ``hard''
corrections as described in \cite{Brignole:2003cm}.
Proton decay is sufficently
and naturally suppressed, even for a rather low cutoff scale.
The lightest particle of the O'Raifeartaigh sector (the LOP) is stable,
and is an interesting cold dark matter candidate.

\subsection*{Acknowledgments}
We thank Bogdan Dobrescu for useful discussions.
This research was supported by the U.S.~Department of Energy
Grants DE-AC02-76CHO3000 and DE-FG02-90ER40560.

\appendix
\section{Appendix}

\subsection{ Left-chiral superfields with their gauge
quantum numbers and O'charge}
\begin{center}
\begin{tabular}[c]{|c|c|c|c|c|c|}
\hline
$\Phi$ & $SU(3)_C$ & $SU(2)_L$ & $U(1)_Y$ & $\tilde{U}(1)$ & O'charge \\
\hline $Q_i$ & 3 & 2 & 1/6 & 1 & 0\\
$U_i^c$ & $\bar{3}$ & 1 & -2/3 & 1 & 0\\
$D_i^c$ & $\bar{3}$ & 1 & 1/3 & 1 & 0\\
$L_i$ & 1 & 2 & -1/2 & 1 & 0\\
$E_i^c$ & 1 & 1 & 1 & 1 & 0\\
$\mathcal{H}$ & 1 & 2 & -1/2 & -2 & 0\\
$\mathcal{H'}$ & 1 & 2 & 1/2 & -2 & 0\\
$K$ & 8 & 1 & 0 & -2 & 1\\
$T$ & 1 & 3 & 0 & -2 & 3\\
$J_j$ & 1 & 1 & 1 & -2 & 1\\
$J_j^c$ & 1 & 1 & -1 & -2 & 1\\
$X_1$ & 1 & 1 & 0 & 4 & 2\\
$X_2$ & 1 & 1 & 0 & -4 & 2\\
$X_3$ & 1 & 1 & 0 & 4 & 2\\
$S$ & 1 & 1 & 0 & 4 & 2\\
$N$ & 1 & 1 & 0 & -4 & 2\\
$P$ & 1 & 1 & 0 & 4 & 2\\
$R_k$ & 1 & 1 & 0 & -2 & 1\\
$O_l$ & 1 & 1 & 0 & 1 & 2\\
$V$ & 1 & 1 & 0 & 4 & 0\\
$Y$ & 1 & 1 & 0 & 0 & 0\\
$\phi$ & 1 & 1 & 0 & 0 & 0\\
\hline
\end{tabular}

where $i=1,2,3$; $j=1,2$; $k=1,..,11$; $l=1,2,3$.
\end{center}

\subsection{Minimization of the Scalar Potential}

The scalar potential consists of two sets of terms - one arising
from the superpotential and the other arising from the $D$ terms.
Thus the scalar potential $V$ can be written as $V=V_W+V_D$. From
the squark and slepton dependence on the scalar potential, it is
straightforward to show that all of them have zero vevs. The
expression for $V_W$, omitting the squarks and sleptons, is given
by:

{\setlength\arraycolsep{2pt}
\begin{eqnarray}
V_W &=& 4{\lambda}_{K}^2|K|^2|X_1|^2+4{\lambda}_{T}^2|T|^2|X_1|^2+
4{\lambda}_{R}^2\sum_{i}^{11}|R_i|^2|X_1|^2
+\sum_{i}^{2}|J_i|^2|{\lambda}_{J}X_1+{\lambda}_{P}P|^2+
\nonumber \\
& &\sum_{i}^{2}|J_i^c|^2|{\lambda}_{J}X_1+{\lambda}_{P}P|^2+
M_1^2|X_2|^2+f^2|X_1X_2-{\mu}^2|^2+|({\mu}'
+\bar{g}\phi)\,N-g_H\,\mathcal{H}\mathcal{H'}|^2+
\nonumber \\
& &
|{\lambda}_{K}K^2+{\lambda}_{T}T^2+\sum_{i}^{2}{\lambda}_{J}J_iJ_i^c+
{\lambda}_{R}\sum_{i}^{11}R_i^2+(fY+M_2)X_2|^2+|\bar{M}N+{\lambda}_P
J_iJ_i^c|^2+\nonumber\\ &
&|M_1X_3+(M_2+f\,Y)X_1|^2+|({\mu}'+\bar{g}\phi)S+\bar{M}P|^2+
|\beta{\phi}^2+m\phi-{\kappa}^2+\bar{g}S\,N|^2+ \nonumber
\\ & &g_H^2\,|S|^2(|H_u^+|^2+|H_d^-|^2+|H_u^0|^2+|H_d^0|^2)
\end{eqnarray}}
The total potential obtained by the sum of $V_D$ and $V_W$ can be
minimized by solving the following set of equations:
{\setlength\arraycolsep{2pt}
\begin{eqnarray}
\frac{\partial V}{\partial K^*}&=&\langle K \rangle
[{\lambda}_{K}^2|K|^2+4{\lambda}_{K}^2|X_1|^2-2{\tilde{g}}^2
\langle \tilde{D} \rangle ]+
 \lambda_K \langle K^* \rangle[\lambda_T T^2+\nonumber\\
& &\lambda_J\sum_i^2J_iJ_i^c+\lambda_R\sum_{i}^{11}R_i^2+(fY+M_2)X_2]=0 \nonumber\\
\frac{\partial V}{\partial T^*}&=& \langle T \rangle
[{\lambda}_{T}^2|T|^2+4{\lambda}_{T}^2|X_1|^2-2{\tilde{g}}^2
\langle \tilde{D} \rangle+ g^2 \langle D_2 \rangle
T^{(2)}_{adj}]+\lambda_T \langle T^*\rangle[{\lambda}_K
K^2+\nonumber\\
&
&\lambda_J\sum_i^2J_iJ_i^c+\lambda_R\sum_{i}^{11}R_i^2+(fY+M_2)X_2]=0 \nonumber\\
\frac{\partial V}{\partial R_i^*}&=&\langle R_i \rangle
[{\lambda}_{R}^2\sum_{i}^{11}|R_i|^2+4{\lambda}_{R}^2|X_1|^2-2{\tilde{g}}^2
\langle \tilde{D} \rangle ]+
 \lambda_R \langle R_i^* \rangle[\lambda_T T^2+\nonumber\\
& & \lambda_J \sum_i^2J_i J_i^c+\lambda_K
K^2+(f\,Y+M_2)X_2]=0 \nonumber\\
\frac{\partial V}{\partial J_i^*}&=&\langle J_i \rangle
[2{\lambda}_{J}^2|J_i^c|^2+{\lambda}_{J}^2|X_1+P|^2+g_y^2 \langle
D_y \rangle-2 \tilde{g}^2 \langle \tilde{D} \rangle ]+
 \lambda_J \langle J_i^{c*} \rangle[\lambda_T T^2+ \nonumber\\
 & &\lambda_K K^2+\lambda_R\sum_{i}^{11}R_i^2+(fY+M_2)X_2+\bar{M}N]=0 \nonumber\\
 \frac{\partial V}{\partial J_i^{c*}}&=&\langle J_i^c \rangle
[{\lambda}_{J}^2|J_i|^2+{\lambda}_{J}^2|X_1+P|^2-g_y^2 \langle D_y
\rangle-2 \tilde{g}^2 \langle \tilde{D} \rangle ]+
 \lambda_J \langle J_i^{*} \rangle[\lambda_T T^2+\nonumber\\
 & & \lambda_K K^2+\lambda_J
J_jJ_j^c+\lambda_R\sum_{i}^{11}R_i^2+(fY+M_2)X_2+\bar{M}N]=0 \nonumber\\
\frac{\partial V}{\partial X_3^*}&=&\langle X_3 \rangle
[M_1^2+4\tilde{g}^2\langle \tilde{D} \rangle]+\langle X_1 \rangle
[M_1(M_2+f Y)] = 0 \nonumber\\
\frac{\partial V}{\partial Y^*}&=&f(f\langle Y \rangle +M_2)(|
X_1|^2+|X_2|^2)+f\langle X_1^* \rangle (M_1
X_3)+\nonumber\\
& &f\langle X_2^* \rangle[\lambda_T T^2+ \lambda_K K^2+\lambda_J
\sum_i^2J_iJ_i^c+\lambda_R\sum_{i=1}^{11}R_i^2]=0
\end{eqnarray}}
{\setlength\arraycolsep{2pt}
\begin{eqnarray}
\frac{\partial V}{\partial X_1^*}&=&\langle X_1 \rangle
[f^2|X_2|^2+4\tilde{g}^2\langle \tilde{D}
\rangle+4{\lambda}_K^2|K|^2+4{\lambda}_T^2|T|^2+4\lambda_R\sum_{i}^{11}R_i^2+|f\,Y+M_2|^2+\nonumber\\
&
&{\lambda}_J^2|J_i|^2+{\lambda}_J^2|J_i^c|^2]+{\lambda}_J^2(|J_i|^2+|J_i^c|^2)\langle
P \rangle + (f\langle Y^* \rangle + M_2)\,M_1\langle X_3 \rangle
-f^2{\mu}^2\langle X_2^* \rangle=0 \nonumber\\
\frac{\partial V}{\partial X_2^*}&=&\langle X_2 \rangle
[f^2|X_1|^2+M_1^2-4\tilde{g}^2\langle \tilde{D} \rangle]-
f^2{\mu}^2\langle X_1^* \rangle + \nonumber\\
& & (f\langle Y^* \rangle +M_2)[\lambda_T T^2+ \lambda_K
K^2+\lambda_J
\sum_i^2J_iJ_i^c+\lambda_R\sum_{i=1}^{11}R_i^2]=0 \nonumber\\
\frac{\partial V}{\partial S^*}&=&\langle S \rangle
[{\bar{g}}^2|N|^2+4\tilde{g}^2\langle \tilde{D}
\rangle+g_H^2\,h^2+|{\mu}'+\bar{g}\phi|^2]+
\bar{g} \langle N^* \rangle(\beta{\phi}^2+m \phi
-{\kappa}^2)+({\mu}'+\bar{g} {\phi}^*)\bar{M}\langle P \rangle=0 \nonumber\\
\frac{\partial V}{\partial N^*}&=&\langle N \rangle
[{\bar{g}}^2|S|^2+{\bar{M}}^2-4\tilde{g}^2\langle \tilde{D}
\rangle+|{\mu}'+\bar{g}\phi|^2]+ \bar{g} \langle S^*
\rangle(\beta{\phi}^2+m \phi-{\kappa}^2)+\bar{M}\langle J_i
\rangle \langle J_i^c
\rangle-\nonumber\\
& &g_H({\mu}'+\bar{g} \langle {\phi}^* \rangle)(H_u^+H_d^- - H_u^0H_d^0) =0 \nonumber\\
\frac{\partial V}{\partial P^*}&=&\langle P \rangle
[{\bar{M}}^2+4\tilde{g}^2\langle \tilde{D}
\rangle+{\lambda}_J^2(|J_i|^2+|J_i^c|^2)]+{\lambda}_J^2\langle X_1
\rangle(|J_i|^2+|J_i^c|^2)+
({\mu}'+\bar{g} {\phi}^*)\bar{M}\langle S \rangle =0 \nonumber\\
\frac{\partial V}{\partial {\phi}^*}&=& \langle \phi \rangle
(2{\beta}^2|\phi|^2+m^2)+2\langle {\phi}^* \rangle
\beta(m\phi-{\kappa}^2+\bar{g}SN)+m(\beta{\phi}^2-{\kappa}^2+\bar{g}SN)\nonumber \\
& &- \bar{g}g_H\langle N^* \rangle(H_u^+H_d^- - H_u^0H_d^0)
+\bar{g}\bar{M}S^*\langle P
\rangle+\bar{g}({\mu}'+\bar{g}\phi)(|S|^2+|N|^2)=0\label{eq:phi}\nonumber
\end{eqnarray}}
Expanding $\tilde{g}^2\langle \tilde{D}\rangle$ in terms of
fields, and plugging it in the equations for
$H_u^+,\;H_d^-,\;H_u^0\;\&\;H_d^0$, we get:

{\setlength\arraycolsep{2pt}
\begin{eqnarray}
\frac{\partial V}{\partial H_u^{+*}}&=&\langle H_u^+ \rangle
[g_H^2|S|^2+|H_u^+|^2(\frac{1}{4}(g^2+g_y^2)+4\tilde{g}^2)+
|H_u^0|^2(\frac{1}{4}(g^2+g_y^2)+4\tilde{g}^2)\nonumber \\
& &+|H_d^0|^2(\frac{g^2}{2}+4\tilde{g}^2-\frac{1}{4}(g^2+g_y^2))
+|H_d^-|^2(g_H^2+4\tilde{g}^2-\frac{1}{4}(g^2+g_y^2))-\nonumber\\
&&\hspace{-15pt}
8\tilde{g}^2(X_1^2-X_2^2+S^2-N^2+\frac{\tilde{\xi}}{4})+\frac{g_y^2\xi}{2}]
+\langle
H_d^{-*}\rangle[(\frac{g^2}{2}-g_H^2)H_u^0H_d^0-g_H({\mu}'+\bar{g}\phi)N]=0\\
\frac{\partial V}{\partial H_d^{-*}}&=&\langle H_d^- \rangle
[g_H^2|S|^2+|H_d^-|^2(\frac{1}{4}(g^2+g_y^2)+4\tilde{g}^2)+
|H_d^0|^2(\frac{1}{4}(g^2+g_y^2)+4\tilde{g}^2)\nonumber \\
& &+|H_u^0|^2(\frac{g^2}{2}+4\tilde{g}^2-\frac{1}{4}(g^2+g_y^2))
+|H_u^+|^2(g_H^2+4\tilde{g}^2-\frac{1}{4}(g^2+g_y^2))-\nonumber\\
&&\hspace{-15pt}
8\tilde{g}^2(X_1^2-X_2^2+S^2-N^2+\frac{\tilde{\xi}}{4})-\frac{g_y^2\xi}{2}]
+\langle
H_u^{+*}\rangle[(\frac{g^2}{2}-g_H^2)H_u^0H_d^0-g_H({\mu}'+\bar{g}\phi)N]=0\\
\frac{\partial V}{\partial H_u^{0*}}&=&\langle H_u^0 \rangle
[g_H^2|S|^2+|H_u^+|^2(\frac{1}{4}(g^2+g_y^2)+4\tilde{g}^2)+
|H_u^0|^2(\frac{1}{4}(g^2+g_y^2)+4\tilde{g}^2)\nonumber \\
& &+|H_d^0|^2(g_H^2+4\tilde{g}^2-\frac{1}{4}(g^2+g_y^2))
+|H_d^-|^2(\frac{g^2}{2}+4\tilde{g}^2-\frac{1}{4}(g^2+g_y^2))-8\tilde{g}^2\nonumber\\
&&\hspace{-15pt}
(X_1^2-X_2^2+S^2-N^2+\frac{\tilde{\xi}}{4})+\frac{g_y^2\xi}{2}]
+\langle H_d^{0*}\rangle[(\frac{g^2}{2}-g_H^2)\langle H_u^+
\rangle \langle H_d^- \rangle+g_H({\mu}'+\bar{g}\phi)N]=0
\label{eq:hu}
\end{eqnarray}}

{\setlength\arraycolsep{2pt}
\begin{eqnarray}
\frac{\partial V}{\partial H_d^{0*}}&=&\langle H_d^0 \rangle
[g_H^2|S|^2+|H_d^-|^2(\frac{1}{4}(g^2+g_y^2)+4\tilde{g}^2)+
|H_d^0|^2(\frac{1}{4}(g^2+g_y^2)+4\tilde{g}^2)\nonumber \\
& &+|H_u^0|^2(g_H^2+4\tilde{g}^2-\frac{1}{4}(g^2+g_y^2))
+|H_u^+|^2(\frac{g^2}{2}+4\tilde{g}^2-\frac{1}{4}(g^2+g_y^2))-8\tilde{g}^2\nonumber\\
&&\hspace{-15pt}
(X_1^2-X_2^2+S^2-N^2+\frac{\tilde{\xi}}{4})-\frac{g_y^2\xi}{2}]
+\langle H_u^{0*}\rangle[(\frac{g^2}{2}-g_H^2)\langle H_u^+
\rangle \langle H_d^-
\rangle+g_H({\mu}'+\bar{g}\phi)N]=0\label{eq:hd}
\end{eqnarray}}
\hspace{8cm}-----------------\\ \\
 The solution to the above
equations is given by: {\setlength\arraycolsep{2pt}
\begin{eqnarray}
\langle K \rangle &=& \langle T \rangle =\langle R_i \rangle =
\langle J_i \rangle = \langle J_i^c \rangle = \langle X_3 \rangle
= \langle P \rangle = \langle H_u^+ \rangle = \langle H_d^- \rangle = 0;\nonumber \\
\langle X_1 \rangle \equiv V_1 \neq 0;\;\langle X_2 \rangle
&\equiv& V_2 \neq 0;\;\langle S \rangle \equiv S \neq 0;\langle N
\rangle \equiv N \neq 0;\; \langle H_u^0 \rangle \equiv h_u \neq
0;\;\langle H_d^0 \rangle
\equiv h_d \neq 0; \nonumber\\
\langle Y \rangle &=& -\frac{M_2}{f};\;\langle \phi \rangle =
-\frac{{\mu}'}{\bar{g}} \nonumber
\end{eqnarray}}
provided the following condition is satisfied:
\begin{equation}
(\frac{g^2}{2}-g_H^2)>0 \label{eq:five}
\end{equation}

\noindent The equations for $H_u^0$ and $H_d^0$ boil down to:

{\setlength\arraycolsep{2pt}
\begin{eqnarray}
(g_H^2+4\tilde{g}^2-\frac{g^2+g_y^2}{4})h_d^2
+(\frac{g^2+g_y^2}{4}+4\tilde{g}^2)h_u^2+ \frac{1}{2}g_y^2\xi-
8\tilde{g}^2(V_1^2-V_2^2+S^2-N^2+\frac{\tilde{\xi}}{4}) =
0\label{eq:hu0} \\
(g_H^2+4\tilde{g}^2-\frac{g^2+g_y^2}{4})h_u^2
+(\frac{g^2+g_y^2}{4}+4\tilde{g}^2)
h_d^2-\frac{1}{2}g_y^2\xi-8\tilde{g}^2(V_1^2-V_2^2+S^2-N^2+\frac{\tilde{\xi}}{4})
= 0\label{eq:hd0}
\end{eqnarray}}
\noindent which is the same as in (\ref{eq:six}) and
(\ref{eq:seven}). The solution is given by:

\begin{displaymath}
\left \{
\begin{array}{c}
h_d^2\\
\\
h_u^2\\
\end{array}\right. =
\frac{8{\tilde{g}}^2(V_1^2-V_2^2+S^2-N^2+\tilde{\xi}/4)-\,g_H^2\,S^2}
{g_H^2+8{\tilde{g}}^2} \pm \frac{g_y^2 \xi}{g^2+g_y^2-2g_H^2}
\end{displaymath}
\begin{equation}
h_d^2-h_u^2=h^2\cos(2\beta)=\frac{g_y^2\xi}{[(g^2+g_y^2)/2-g_H^2]}
\end{equation}

\noindent We see that the Fayet-Iliopoulos term for $U(1)_y$ -
$\xi$, determines $\tan{\beta}$ and in combination with $V_1$, the
Z boson - B' boson mixing (\ref{eq:eleven}).

\noindent Similarly, with the above vacuum, the equations for
$V_1,V_2,S\,\&\,N$ boil down to:

\begin{eqnarray}
S[{\bar{g}}^2N^2+4{\tilde{g}}^2\langle \tilde{D} \rangle +
g_H^2\,h^2]&=&-\bar{g} N
(\beta{\phi}^2+m\phi-{\kappa}^2)\label{eq:s}\\
N[{\bar{g}}^2S^2-4{\tilde{g}}^2\langle \tilde{D} \rangle +
{\bar{M}}^2]&=&-\bar{g} S
(\beta{\phi}^2+m\phi-{\kappa}^2)\label{eq:n}\\
4{\tilde{g}}^2\langle \tilde{D} \rangle + f^2\,V_2^2 &=&
f^2\,{\mu}^2(\frac{V_2}{V_1})\label{eq:v1}\\
-4{\tilde{g}}^2\langle \tilde{D} \rangle + f^2\,V_1^2 + M_1^2&=&
f^2\,{\mu}^2(\frac{V_1}{V_2})\label{eq:v2}
\end{eqnarray}

\noindent which is the same as in (\ref{eq:two}). Now,
${\tilde{g}}^2\langle \tilde{D} \rangle$ is given in terms of
fields by:
\begin{equation}
{\tilde{g}}^2\langle \tilde{D} \rangle =
-2{\tilde{g}}^2\,h^2+4{\tilde{g}}^2(V_1^2-V_2^2+S^2-N^2+\tilde{\xi}/4)
\label{eq:dtwiddle}
\end{equation}

\noindent As an illustration, we can fix the
vevs $V_1=V_2=S=20\;\mathrm{TeV}$ and show that all other vevs and
dimensionful parameters are also of the same scale.

\noindent From
(\ref{eq:hu0}),(\ref{eq:hd0}),(\ref{eq:dtwiddle}),(\ref{eq:s}),(\ref{eq:n}),(\ref{eq:v1})
and (\ref{eq:v2}), we get :
\begin{equation}
4{\tilde{g}}^2\langle \tilde{D} \rangle =
g_H^2(2S^2+h^2)=\frac{M_1^2}{2}=\frac{{\bar{M}}^2N^2-g_H^2h^2S^2}{S^2+N^2}
\end{equation}
Therefore $ N^2 =
\frac{2g_H^2S^2(S^2+h^2)}{({\bar{M}}^2-2g_H^2S^2-g_H^2h^2)}$.
Choosing ${\bar{M}}^2=g_H^2(4S^2+h^2)$ for convenience gives us:
\begin{eqnarray}
N^2=S^2+h^2;\;{\mu}^2=S^2(1+\frac{2g_H^2}{f^2}+\frac{g_H^2\,h^2}{f^2S^2});\;M_1^2=g_H^2(4S^2+2h^2)
\end{eqnarray}

\noindent Also, from (\ref{eq:s}),(\ref{eq:n}) and (\ref{eq:phi}),
we get :
\begin{eqnarray}
\beta{\langle \phi \rangle}^2+m \langle \phi \rangle - {\kappa}^2
+\bar{g}S\,N&=&
-\frac{S\,N}{\bar{g}(S^2+N^2)}[{\bar{M}}^2+g_H^2h^2]\\
\Rightarrow  m &=&
\frac{2\beta}{\bar{g}}{\mu}'+(\frac{\bar{g}g_H\sin{2\beta}}{2\mathcal{A}})(\frac{h^2}{S})\\
\mathrm{where}\;\; \mathcal{A}&=&
\frac{({\bar{M}}^2+g_H^2h^2)}{\bar{g}(S^2+N^2)}
\end{eqnarray}

\noindent Putting in the numbers, we obtain the values of
dimensionful parameters, as in (\ref{eq:eight}). Finally, the
Fayet-Iliopoulos terms for $\tilde{U}(1)$ are given by:

\begin{equation}
\tilde{g}^2\,\tilde{\xi}=
(\frac{g_H^2}{4}+6\tilde{g}^2)h^2+\frac{g_H^2}{2}S^2
\end{equation}

\newpage


\end{document}